\documentclass[10pt,letterpaper]{article}
\usepackage[letterpaper,margin=1in]{geometry}

\usepackage{amsmath,amssymb}

\usepackage{changepage}

\usepackage{textcomp,marvosym}

\usepackage{cite}

\usepackage{nameref,hyperref}

\usepackage[right]{lineno}

\usepackage[nopatch=eqnum]{microtype}
\DisableLigatures[f]{encoding = *, family = * }

\usepackage[table]{xcolor}

\usepackage{array}

\newcolumntype{+}{!{\vrule width 2pt}}

\newlength\savedwidth

\raggedright
\usepackage[aboveskip=6pt,labelfont=bf,labelsep=period,justification=raggedright,singlelinecheck=off]{caption}

\makeatletter
\renewcommand{\@biblabel}[1]{\quad#1.}
\makeatother

\usepackage{lastpage,fancyhdr,graphicx}
\usepackage{epstopdf}
\usepackage{subfigure}
\usepackage{booktabs} 
\usepackage{physics}
\usepackage{tikz}
\usetikzlibrary{bayesnet}
\usepackage{multirow}

\usepackage{mathtools}
\usepackage{amsthm}

\usepackage[capitalize,noabbrev]{cleveref}

\theoremstyle{plain}

\theoremstyle{definition}

\theoremstyle{remark}

\usepackage[textsize=tiny]{todonotes}

\begin{document}
\vspace*{0.2in}

\begin{flushleft}
{\Large
\textbf\newline{Generative Modeling of Neural Dynamics via Latent Stochastic Differential Equations} 
}
\newline
\\
Ahmed ElGazzar\textsuperscript{1*},
Marcel van Gerven\textsuperscript{1}
\\
\bigskip
\textbf{1} Department of Machine Learning and Neural Computing, Donders Institute for Brain, Cognition and Behaviour, Radboud University, Nijmegen, the Netherlands
\\
\bigskip

%
%





* ahmed.elgazzar@donders.ru.nl

\end{flushleft}
\section*{Abstract}
Understanding how neural circuits implement computation remains a central challenge in neuroscience, requiring models that can unite mechanistic interpretability with the complexity of real neural data. Current approaches face a fundamental trade-off: mechanistic biophysical and phenomenological models are difficult to fit to single-trial data, while flexible machine learning models lack interpretability and theoretical grounding.
We propose a probabilistic framework that addresses this challenge by combining domain-specific mathematical models with neural networks through stochastic differential equations (SDEs). In this framework, physiological recordings are viewed as discrete-time partial observations of an underlying continuous-time stochastic dynamical system that implements computations through its state evolution. We employ coupled SDEs with differentiable drift and diffusion functions and use variational inference to infer system states and parameters. This formulation enables seamless integration of existing mathematical models from the literature, neural networks, or hybrid combinations to learn and compare different modeling approaches. We demonstrate this framework by developing a generative model that combines coupled oscillators with neural networks to capture latent population dynamics from single-cell recordings. Evaluation across three neuroscience datasets spanning different species, brain regions, and behavioral tasks shows that these hybrid models achieve competitive performance in predicting stimulus-evoked neural and behavioral responses compared to sophisticated black-box approaches, while requiring an order of magnitude fewer parameters, providing principled uncertainty estimates, and offering interpretable mathematical descriptions of neural dynamics.\footnote{The code is publicly available at \url{https://github.com/elgazzarr/LdX}}

\section*{Introduction}\label{intro}
Unlike many scientific fields, neuroscience lacks a unified theoretical framework that bridges micro-scale mechanisms with macro-scale observations. That is, while the biophysical processes underlying individual neuron spiking have been well-understood for decades~\cite{hodgkin1952quantitative}, the computational principles governing collective neural activity that give rise to cognition and behavior remain elusive.

Dynamical systems theory provides a promising framework to address this gap~\cite{Rabinovich2006Dynamicala,durstewitz2023reconstructing}. Within this formalism, the brain is viewed as a complex dynamical system,  where neural activity patterns evolve through a high-dimensional state space according to well-defined governing equations, providing a mathematical language for understanding how distributed neural circuits implement computation and generate behavior~\cite{van1998dynamical, izhikevich2007dynamical, deco2008dynamic, breakspear2017dynamic, favela2021dynamical}. This perspective has led to diverse mathematical models, ranging from mean-field approximations of biophysical models that preserve mechanistic interpretability \cite{wilson1972excitatory, jirsa1996field, robinson1997propagation} to phenomenological models inspired by statistical physics that capture emergent collective behavior~\cite{amari1977dynamics, hopfield1982neural, sompolinsky1988chaos}. However, while these models have provided key insights into various neural phenomena~\cite{goldman1995cellular, wang2002probabilistic, fries2005mechanism}, fitting them to single-trial neural data has proven challenging~\cite{urai2022large}.

In recent years, latent variable models have emerged as a powerful alternative. These models represent high-dimensional neural activity in terms of low-dimensional latent states evolving according to Markovian dynamics~\cite{paninski2010new,hurwitz2021building}.
Neural computations are thus assumed to be implemented through dynamic motifs in a low-dimensional latent space (e.g., attractors, oscillations, bifurcations)~\cite{vyas2020computation,khona2022attractor, durstewitz2023reconstructing}.
The popularity of this approach stems from converging empirical evidence that neural population activity during simple tasks lives on a low-dimensional manifold~\cite{gao2015simplicity}. Coupled with ongoing developments in machine learning~\cite{kingma2013auto, fabius2014variational, rubanova2019latent}, latent variable models are becoming a standard tool in systems and computational neuroscience~\cite{chen2018neural, glaser2020recurrent, hurwitz2021targeted, zhou2020learning, kim2021inferring}.

However, selecting an appropriate model for representing latent neural dynamics presents a significant challenge. The model must be expressive enough to capture complex non-linear dynamics, supporting flexible neural computations, while remaining amenable to interpretation and analysis to be useful for hypothesis generation and testing. Furthermore, model predictions should account for various sources of uncertainty typically present in such modeling endeavors, such as measurement noise, process noise, and model uncertainty. Finally, to utilize the inferred model for online applications, as in brain-computer interfaces, there should be an efficient tractable method for sampling from the model in real time.

We posit that recent developments in scientific machine learning -- where domain-specific models, typically expressed using differential equations, are combined with neural networks and scalable optimization techniques -- provide an untapped potential to address these challenges~\cite{raissi2019physics,rackauckas2020universal,lai2021structural,karniadakis2021physics}. Additionally, ongoing innovations in training (neural) stochastic differential equations (SDEs)~\cite{tzen2019neural,li2020scalable,zeng2023latent, course2024amortized}, provide an opportunity to develop scalable probabilistic models in neuroscience, a field wherein the systems in question are inherently stochastic~\cite{laing2009stochastic,rolls2010noisy} and the existing measurement tools only provide a coarse proxy of the system under study~\cite{stevenson2011advances, urai2022large}.

To make these advances in scientific machine learning available to the neuroscience community, we developed a general framework for modeling neural dynamics using latent SDEs with differentiable state- and input-dependent drift and diffusion functions. Within this framework, we introduce a specific generative model, combining coupled oscillators with neural networks, to capture neural population dynamics. We evaluate our approach across multiple experimental scenarios, demonstrating its effectiveness and efficiency in predicting stimulus-evoked neural and behavioral responses in both reaching and decision-making tasks. Through systematic comparison of latent ODE- and SDE-based models under varying noise levels and state dimensionality in simulated systems, we highlight the critical importance of modeling process noise; a factor often overlooked in current latent variable approaches. Finally, our results suggest that data-driven coupled oscillators could provide an expressive yet interpretable framework for understanding neural dynamics, while maintaining computational tractability.

\subsection*{Related work}

\paragraph{Latent variable models in neuroscience}
Our work builds on previous latent variable models which have been proposed to infer low-dimensional latent dynamics from single-trial data~\cite{sahani1999latent, yu2008gaussian, macke2011empirical, wu2017gaussian, pandarinath2018inferring, duncker2019, hurwitz2021targeted, schimel2021ilqr, kim2021inferring, kim2023flow}. Most related to our work, is Gaussian process switching linear dynamical system~\cite{hu2024modeling} which leverage SDEs to represent latent neural dynamics extending previous work proposed by Duncker et. al~\cite{duncker2019}. In their work, the authors developed a novel Gaussian process kernel function that defines a smooth, locally linear prior on dynamics, yielding easily interpretable recurrent linear switching dynamical systems while providing posterior uncertainty estimates. Compared to Gaussian process-based methods, our approach provides flexibility by allowing the modeler to choose any dynamical system as the prior, enabling systematic model comparison across a wide range of existing neural models. Additionally, our approach is not limited to a small number of latent state parameters as is often the case with GP-based approaches, enabling greater flexibility in model capacity. Finally, unlike~\cite{hu2024modeling}, the SDEs employed in our approach are not restricted to additive process noise, but can model multiplicative (state- and input-dependent) process noise, enabling greater flexibility and model expressivity. Other relevant recent work extends low-rank RNNs to a stochastic setting with additive process noise~\cite{pals2024inferring}. Here, the authors use a specific form of the RNN which enables finding fixed-points effectively after training. 

\paragraph{Scalable inference of latent SDEs}
Inferring the state and learning the parameters of a latent SDE from (high-dimensional) noisy observations is a challenging problem which appears throughout science and engineering. With the advent of neural differential equations~\cite{chen2018neural, kidger2022neural}, there has been a growing interest in developing scalable methods for fitting latent neural SDEs to data. Seminal work by~\cite{tzen2019neural, tzen2019theoretical} established the theoretical foundation for training generative neural SDEs via variational inference through the lens of stochastic control. Subsequent research has focused on addressing key computational challenges, including improved memory efficiency and extension to the time domain~\cite{li2020scalable}, enhanced numerical stability~\cite{zeng2023latent, oh2024stable}, and computational efficiency~\cite{course2024amortized}. Our work builds upon the formulation by~\cite{li2020scalable}, extending it to controlled and multi-modal settings in neuroscience. 
This specific formulation enables preserving the expressivity of the generative SDE through state- and input-dependent diffusion on arbitrary spaces.

\paragraph{Modeling neural dynamics with coupled oscillators}
Coupled oscillator models have a rich history in theoretical neuroscience, dating back to seminal work by~\cite{kuramoto1975self} and~\cite{wilson1972excitatory}. These models have proven particularly effective in capturing rhythmic neural activity and investigating synchronization phenomena in neural circuits~\cite{breakspear2010generative, ermentrout2010mathematical}. Of particular relevance to our work are modern approaches that combine oscillator models with machine learning techniques~\cite{ abrevaya2021learning, abrevaya2023effective, bandyopadhyay2023phenomenological} to learn dynamical systems that can reproduce complex neural trajectories while maintaining interpretable structure. Our approach builds upon these foundations by introducing a flexible SDE framework that can naturally capture oscillatory dynamics while accounting for process noise and external inputs. This bridges the gap between mechanistic oscillator models and data-driven approaches, allowing for both accurate trajectory prediction and meaningful scientific interpretation.

\section*{Methods}

\paragraph{Notation} 
Let $x\colon [0, \tau] \to \mathbb{R}^{d_x}$ denote a continuous path of latent neural states, which we observe through neural recordings (e.g., spike counts, fMRI time-series) denoted by $\mathcal{Y} = \{{y}_{i}\}_{i=1}^{T}$ with ${y}_{i} \in \mathbb{R}^{d_y}$ is observed at time $t_i \in [0, \tau]$. 
Additionally, we may observe behavioral responses (e.g., movement trajectories, choice responses) represented as $\mathcal{B} = \{{b}_{i}\}_{i=1}^{T}$ with ${b}_{i} \in \mathbb{R}^{d_b}$. In some settings, these observations are driven by external stimuli or task variables (e.g., visual stimuli, auditory cues) denoted by $\mathcal{V} = \{{v}_{i}\}_{i=1}^{T}$ with ${v}_{i} \in \mathbb{R}^{d_v}$. A single trial is thus defined as $\mathcal{S} = (\mathcal{Y}, \mathcal{B}, \mathcal{V})$, where $\mathcal{Y}$, $\mathcal{B}$ and/or $\mathcal{V}$ could be (partially) missing. Given a dataset $\mathcal{D} = \{\mathcal{S}_k\}_{k=1}^K$ consisting of $K$ trials, our objectives are to: (1) infer the underlying neural dynamics that generated the observations, represented by the posterior distribution $p(x \mid \mathcal{D})$, (2) learn a generative model capable of predicting stimulus-evoked responses, represented by the joint distribution $p(\mathcal{Y}, \mathcal{B} \mid \mathcal{V})$, and (3) extract interpretable dynamical features that provide insights into neural computations.

\subsection*{Generative model}
We model latent neural dynamics $x$ as a continuous-time stochastic process, specified as the solution to the following Itô SDE:
\begin{equation}
   \dd x(t) = \mu_{\theta}(x(t), u(t))\dd t + \sigma_{\theta}(x(t), u(t)) \dd W(t)
\end{equation}
with $x_0 \sim \mathcal{P}_0$, where $\mathcal{P}_0$ is the probability distribution of the initial condition $x_0 \coloneq x(0)$, and $u(t) = \sum_{i=1}^T \eta_\theta({v}_{i}) \psi_i(t)$ is an interpolated encoded representation of $\mathcal{V}$. Here, $\{\psi_i\}_{i=1}^T$ are basis functions chosen for the desired interpolation scheme and  $\eta_{\theta}\colon \mathbb{R}^{d_v} \to \mathbb{R}^{d_u}$ is an input encoder whose form depends on the input modality. 
The functions $\mu_{\theta}\colon \mathbb{R}^{d_x} \times \mathbb{R}^{d_u} \to \mathbb{R}^{d_x}$ and $\sigma_{\theta}\colon \mathbb{R}^{d_x} \times \mathbb{R}^{d_u} \to \mathbb{R}^{d_x \times d_W}$ are Lipschitz-continuous differentiable functions defining the drift and diffusion coefficients of the SDE, respectively. Finally, $W\colon [0,\tau] \to \mathbb{R}^{d_W}$ denotes a standard $d_W$-dimensional Wiener process.

This SDE induces a path measure on the space of continuous paths which implicitly define the distribution $p(x \mid \mathcal{V})$. Samples from this distribution for $t > 0$ are obtained by solving the SDE with different realizations of $x_0$ and $W$:
\begin{equation}
    x(t) = x_0 + \int_0^t \mu_{\theta}(x(s), u(s)) \dd s + \int_0^t \sigma_{\theta}(x(s), u(s)) \dd W(s) \,.
\end{equation}
The continuous-time latent states $x_i = x(t_i)$ are mapped to discrete-time observations through:
\begin{equation}
   {y}_{i} \sim p(\cdot \mid \lambda_\theta(x_i)), \quad {b}_{i} \sim p(\cdot \mid \rho_\theta(x_i))
\end{equation}
where $p(\cdot \mid \lambda_\theta(x))$ and $p(\cdot \mid \rho_\theta(x))$ denote the observation models (e.g., Poisson model for discrete measurements, Gaussian model for continuous measurements) for neural and behavioral data, respectively, with $\lambda_\theta$  and $\rho_\theta$ being differentiable functions mapping the latent state to the parameters of these distributions.

\begin{figure*}[t] 
    \centering
    \includegraphics[width=1.0\textwidth]{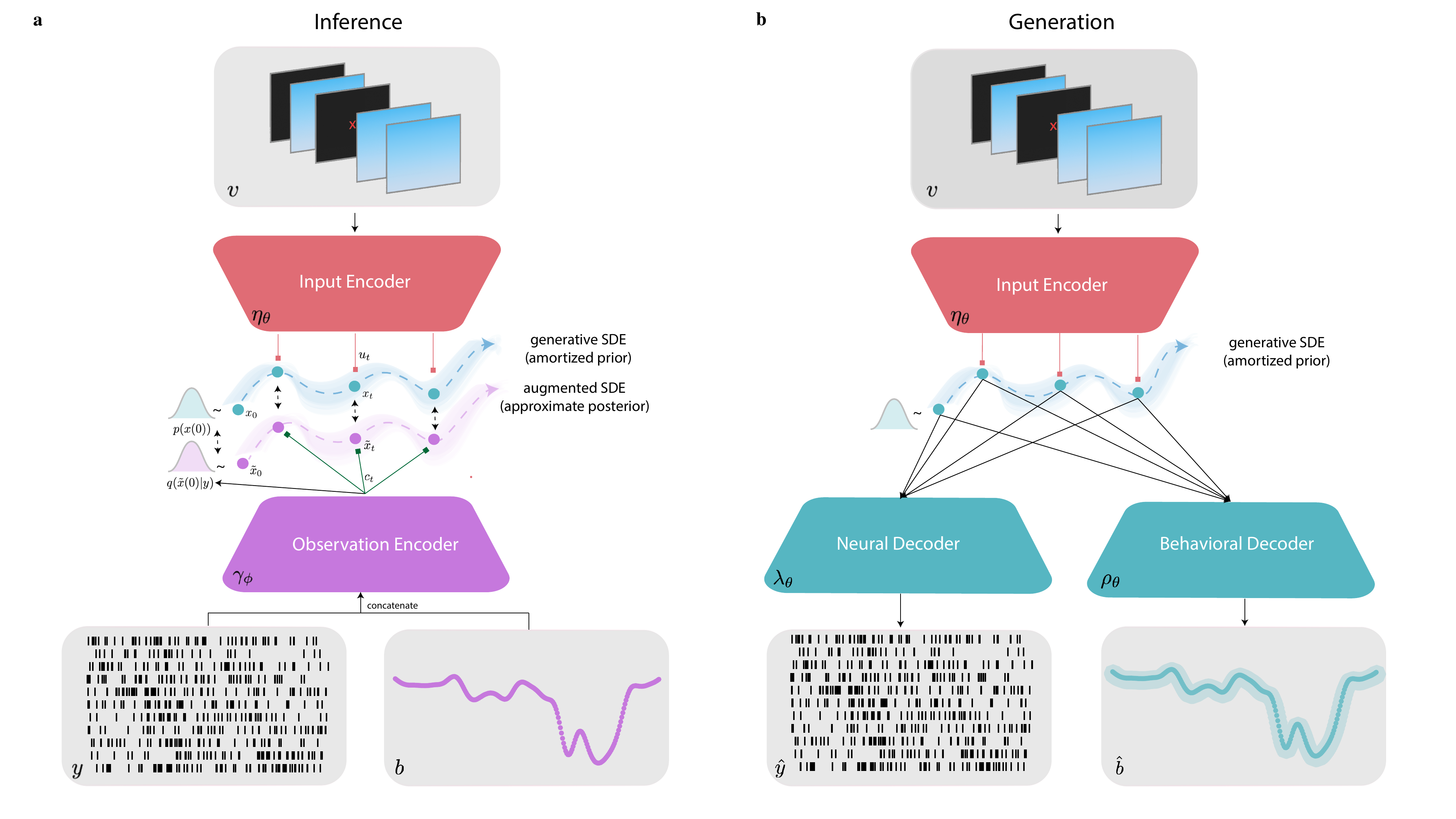}
    \caption{\textbf{Schematic illustration of the framework.} \textbf{a} The inference model. Experimental stimuli $v$ are processed by an input encoder ($\eta_{\theta}$) to produce a continuous-time input function $u$. Neural observations $y$ and behavioral observations $b$ are concatenated and processed by an observation encoder ($\gamma_{\phi}$) to produce a continuous-time context $c$ and the distribution parameters of an initial density $q(\tilde{x}_0 \mid y)$. An augmented SDE, serving as the approximate posterior, takes an initial condition $\tilde{x}_0$, $u_t$, and $c_t$ to infer latent trajectories $\tilde{x}_t$ at time $t > 0$. This augmented SDE is trained to match the generative SDE serving as our prior, which models latent trajectories $x_t$ based on an initial condition $x_0$ and $u_t$. The model parameters $\theta$ and variational parameters $\phi$ are learned by maximizing the evidence lower bound using variational inference.
    \textbf{b} The generation model. After training, given a stimulus, the input encoder ($\eta_{\theta}$) generates and encoded input signal $u$. The learned generative SDE uses $u(t)$ as a control signal and an initial state $x_0$ sampled from $p(x_0)$ to generate latent trajectories $x(t)$ (In the presence of historical observations, we sample the initial state from the approximate posterior conditioned on the observations). These trajectories are then passed to a neural decoder ($\lambda_{\theta}$) to predict neural activity $\hat{y}$ and a behavioral decoder ($\rho_{\theta}$) to predict behavioral responses $\hat{b}$.}
    \label{fig_framework}
\end{figure*}

\subsection*{Model estimation}
The true parameters $\theta$ of our generative model are unknown and need to be learned from data $\mathcal{D}$. Moreover, the exact posterior distribution $p(x \mid \mathcal{D})$ is intractable to compute. Variational inference provides an efficient and scalable approach to tackle this problem. Unlike standard latent variable models, the latent state in our model are realizations of a continuous-time stochastic process.  Thus we need to define a parametric distribution over paths rather than random-valued vectors. 
Figure~\ref{fig_framework} provides an overview of our framework.

Following~\cite{tzen2019neural,tzen2019theoretical,li2020scalable}, we define the approximate posterior distribution $q(x \mid \mathcal{D})$ via an augmented (neural) SDE. Specifically, we define the approximate posterior path $\tilde{x}\colon [0, \tau] \to \mathbb{R}^{d_x}$ as realizations of the following SDE (we omit time indices for ease of notation):
\begin{equation}
    \dd \tilde{x} = \nu_{\phi}(\tilde{x}, u, c) \dd t + \sigma_{\theta}(\tilde{x}, u) \dd W
\end{equation}
where $\tilde{x}_0 \sim \mathcal{Q}_0$ and $c(t) = \sum_{i=1}^T \xi_\phi({y}_{i}, {b}_{i}, {v}_{i}) \psi_i(t)$ is a continuous path encoding the observations which we refer to as context, with $\xi_\phi \colon \mathbb{R}^{d_y \times d_b \times d_v} \to \mathbb{R}^{d_c}$ being a modality-specific observation encoder.
We define the distribution of the initial condition as $\mathcal{Q}_0 = \mathcal{N}(\alpha_{\phi}({c}_{1:n}), \beta_{\phi}({c}_{1:n}))$, where $\alpha_{\phi}$ and $\beta_{\phi}$ are neural networks, and ${c}_{1:n} = \{{c}_1,\ldots, {c}_n\}$ are discrete samples of the continuous-time context $c(t)$.
Here, $\nu_{\phi} \colon \mathbb{R}^{d_x} \times \mathbb{R}^{d_u} \times \mathbb{R}^{d_c}  \to \mathbb{R}^{d_x}$ 
is a neural network representing the drift term of the augmented SDE. 
Note that while the drift term $\nu_{\phi}$ is different from that of the generative model, the diffusion term $\sigma_{\theta}$ is shared with the generative SDE. 

As described in~\cite{tzen2019neural}, this specification enables leveraging Girsanov's theorem~\cite{girsanov1960transforming} to obtain a tractable KL divergence between the generative (prior) and augmented (approximate posterior) distributions:
 \begin{equation}
     D_\text{KL}(\mathcal{Q}_\tau \| \mathcal{P}_\tau) =
     \mathbb{E}_{\tilde{x}} \left[\int_0^\tau \frac{1}{2} \, \left\| \Delta(\tilde{x}, u, c)\right\|^2 \dd t \right]
 \end{equation}
 where
 $
     \Delta(\tilde{x}, u, c) = \sigma_{\theta}(\tilde{x}, u)^{-1}
     \left(\nu_{\phi}(\tilde{x}, u, c) - \mu_{\theta}(\tilde{x}, u)\right)
 $
with $\mathcal{P}_\tau$ and $\mathcal{Q}_\tau$ denoting the measures induced by the generative and augmented SDE, respectively.
Using this property, we can define the following evidence lower bound (ELBO) on the conditional marginal likelihood of the observations:
\begin{equation}
\log p(\mathcal{Y}, \mathcal{B} \mid \mathcal{V})  \geq
 \mathbb{E}_{\tilde{x}} \left[\sum_{i=1}^T \log p({y}_{i}| \tilde{x}_i) + \sum_{i=1}^T \log p({b}_{i}|\tilde{x}_i)\right]
- D_{KL}(\mathcal{Q}_0 \| \mathcal{P}_0) - D_{KL}(\mathcal{Q}_\tau \| \mathcal{P}_\tau)
\end{equation}
The ELBO consists of three main terms: (1) the expected log-likelihood of the observations under the approximate posterior, which includes both neural recordings and behavioral responses, (2) the KL divergence between the initial state distributions, and (3) the path-wise KL divergence between the approximate posterior and prior processes derived via Girsanov's theorem (see Supplementary Information for a detailed derivation of the loss).
The model parameters $\theta$ and variational parameters $\phi$ can be jointly optimized by maximizing this ELBO using stochastic gradient descent. Both the expectation over $\tilde{x}$ and the path-wise integral in the KL divergence term are computed using numerical integration schemes suitable for SDEs (see Supplementary Information for implementation details).

This formulation enables us to simultaneously learn the parameters of the generative model while performing approximate posterior inference over the latent neural trajectories. The learned generative model can then be used for various online and offline downstream tasks, including probabilistic prediction of stimulus-evoked responses and extraction of interpretable dynamical features.

\subsection*{Biophysics-inspired latent SDE models}
So far, we have described a general framework for modeling the latent dynamics of a neural system from empirical observations. This framework allows us to leverage almost arbitrary differentiable functions to model the drift and diffusion terms of an SDE. 
A natural
choice would be to use a neural SDE to act as the generative model because of their universal approximation properties. However, this comes at the cost of efficiency and interpretability.

An alternative option, is to leverage prior models proposed in the literature and use neural networks either to approximate their parameters or to augment their dynamics. Such hybrid models can improve sample-efficiency and  generalization performance, and are easier to interpret in comparison to models that rely only on neural networks~\cite{rackauckas2020universal,elgazzar2024universal}.

In this section, we detail a specific instantiation of the framework which
combines a well-understood dynamical system and neural networks to obtain an expressive, interpretable and parameter-efficient generative model of neural population dynamics. Furthermore, we describe how such models can be linked to neurobehavioral data via the specification of appropriate forward models.

\subsubsection*{Stochastic nonlinear coupled oscillators}
A well-known class of models that are used to study several
biological and physical systems are coupled oscillators.
Coupled oscillators provide a rich yet simplified language for studying complex interactions in systems exhibiting emergent behavior such as synchronization, pattern formation and phase transitions~\cite{winfree1980geometry}.

We focus on a specific instantiation of coupled oscillators~\cite{matthews1991dynamics} described by the following complex-valued ordinary differential equation (ODE):
\begin{equation}
\dd z = ((\alpha + i\omega)z + |z|^2z + \kappa z) \dd t
\end{equation}
where $z \in \mathbb{C}^{d_z}$ represents the position of $d_z$ oscillators in the complex plane, $\alpha \in \mathbb{R}^{d_z}$ the bifurcation parameters,
$\omega \in \mathbb{R}^{d_z}$ denotes the natural frequency, and $\kappa \in \mathbb{R}$ represents all-to-all coupling strength. This formulation captures the essential dynamics of coupled limit-cycle oscillators near a supercritical Hopf bifurcation, where in the weak coupling regime, the term $|z(t)|^2$ determines the limit cycle radius.
Despite its simplicity, this system can exhibit diverse emergent dynamical regimes including frequency locking, amplitude death, and chaos through variations in its parameters and the initial conditions.

Here we propose to use this system as the drift of our generative SDE in our framework.
Specifically, we use the position of the oscillator as the latent state $x(t)$ in the generative SDE.
Additionally, we consider the coupling strength $\kappa$ as a time-varying input-dependent function of the external input $u(t)$.
Incorporating these updates and writing the model in terms of the real and imaginary parts of the oscillator $z(t) = a(t) + b(t)i$, we obtain the following system of SDEs:
\begin{align}
\dd a &=
\bigl[\alpha a - \omega b - (a^2 + b^2)a
 + \kappa(u)a\bigr]\dd t  + \Sigma_{a}(x,u)\dd W \\
\dd b &=
\bigl[\omega a + \alpha b - (a^2 + b^2)b  + \kappa(u)b \bigr] \dd t  + \Sigma_{b}(x,u)\dd W
\end{align}
where $x=(a,b) \in \mathbb{R}^{d_x}$ represents the state of $N = d_x / 2$ coupled oscillators and 
$\kappa$, $\Sigma_{a}$, $\Sigma_{b}$ are neural networks representing the input dependent connectivity and the diffusion terms for the real and imaginary states respectively.
The trainable parameters $\theta$ of the generative SDE are thus the parameters of the neural networks as well as $\alpha$ and $\omega$. We refer to this dynamics model as a coupled-oscillator SDE (CO-SDE).

\subsubsection*{Modeling neurobehavioral observations}

Often, we have access to observational data, which can come in the form of neural activity recorded from multiple brain regions or in the form of behavioral measurements. To inform the latent SDE model, we need to condition the generative model on such observational data. 

Given access to neural activity  ${y} = ({y}^1, \ldots, {y}^J)$ recorded from $J$ brain regions, we can extend
the generative model by factorizing the latent state $x = ({x}^1, \ldots, {x}^J)$, where the number of oscillators $N_j$ can vary depending on the population size and complexity.
Assuming that the neural data comes in the form of population firing rates, we model the observed activity of a population as 
\begin{equation}
   {y}_i^j \sim \text{Poisson}\bigl({r}\bigr)
\end{equation}
where
$
r = \exp(\lambda_{\theta}^j(x^j_i))
$
is a vector of rate parameters which depends on  a population-specific transformation
$\lambda_{\theta}^j \colon \mathbb{R}^{N_j} \to \mathbb{R}^{N_j}$ that maps the latent state to the firing rates of the population.

For behavioral observations, in case of continuous-valued data (e.g. arm position, pupil diameter), we assume a Gaussian measurement model:
\begin{equation}
b_i \sim \mathcal{N}(\rho_{\theta}(x_i), \rho_{\theta}(x_i))
\end{equation}
where $\rho_{\theta} \colon \mathbb{R}^{d_x} \to \mathbb{R}^{d_b}$ is a neural network that maps the latent state to the mean and variance parameters of the observation distribution, respectively. 

Finally, for discrete-valued behavioral data (e.g., choice response), we employ a categorical observation model:
\begin{equation}
b_i \sim \text{Categorical}(\rho_{\theta}(x_i))
\end{equation}

Note that neural and behavioral data can be easily combined by using the described forward models in parallel. This provides a natural solution to the data fusion problem and provides a principled way to integrate multi-modal data in our analyses.

\section*{Results}

\begin{figure*}[!t]  
    \centering
    \includegraphics[width=1.0\textwidth]{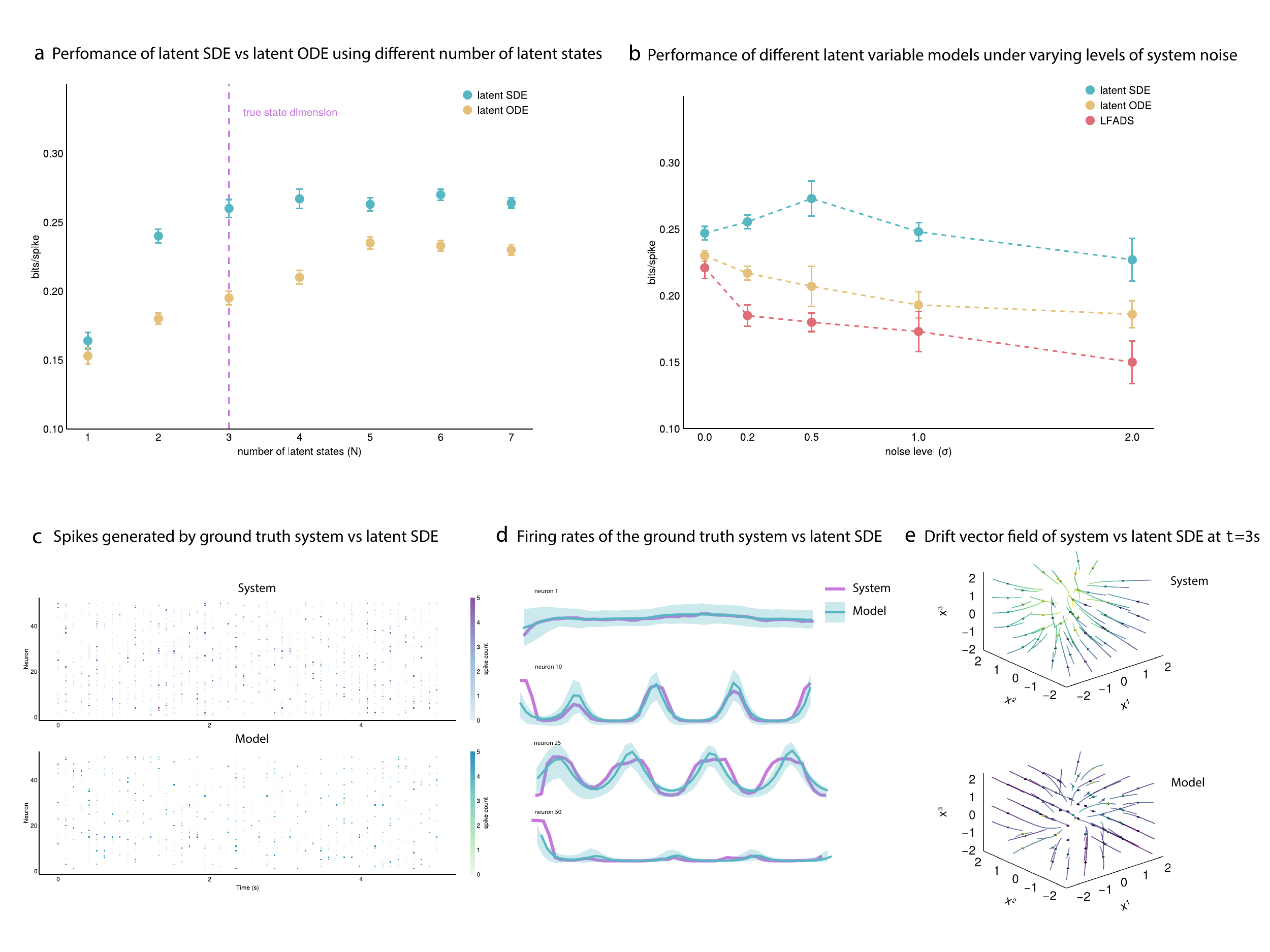}
    \vspace{0.1in}
    \caption{\textbf{Results on a simulated spiking neural system.} \textbf{a} Five-fold cross validation results of comparing  the performance of latent ODE vs latent SDE on fitting simulated data under different number of model latent states. \textbf{b} Five-fold cross validation results of different latent variable models under varying levels of process noise in the ground-truth system. \textbf{c} A sample of the spikes generated via the system vs spikes generated via the latent SDE model in response to the same input. \textbf{d} Samples of underlying firing rates generated via the system versus mean and standard deviation of 30 samples from a trained latent SDE model. \textbf{e} Phase portrait of the ground truth system versus the drift vector field of a trained latent SDE model at one time point under a fixed input.}
    \label{fig_simualtions}
\end{figure*}

\subsection*{Validation on simulated data}
We first sought to evaluate the efficacy of our framework for modeling a partially observed neural system with different levels of process noise.
To this end, we generated spiking data
from the following system:
\begin{align}
x(0) &\sim \mathcal{N}(0,I) \nonumber \\
\dd x &= \delta^{-1} \left(-x + \tanh({A}x + {B}u)\right)\dd t + \sigma \dd W \\
y &\sim \text{Poisson}(\exp({R}x)) \nonumber
\end{align}
where $x(t) \in \mathbb{R}^3$ describes the population average firing rate of three neural populations driven by a sinusoidal input $u(t)$ with randomly sampled frequency and phase. The system matrix 
${A} \in \mathbb{R}^{3 \times 3}$ and input matrix ${B} \in \mathbb{R}^{3 \times 1}$ describe the recurrent connectivity and input dependence, respectively. The time scale of the network dynamics is given by
${\delta} \in \mathbb{R}^3$ and $\sigma \in \mathbb{R}$ describes the magnitude of process noise in the system. The system output is given by $y(t) \in \mathbb{N}_0^{50}$ representing the observed spike counts of 50 neurons generated by sampling from an inhomogeneous Poisson process with rates computed by affine transformation of the population average activity via the output matrix $R \in  \mathbb{R}^{50 \times 3}$. 

We fit our latent SDE model on the spiking data by maximizing the Poisson-log likelihood of the observations given predicted rate and evaluated  performance using bits per spike (bps), a metric that quantifies how well the predicted firing rates explain the observed spike patterns~\cite{pei2021neural}.
For our experiments, we varied the noise level $\sigma \in \{0.0, 0.5, 1.0, 2.0\}$ to test the robustness of the model. We used a latent neural SDE with varying sizes of latent dimension and compared against two baselines: a latent ODE~\cite{rubanova2019latent} and LFADS~\cite{pandarinath2018inferring} as these have gained popularity in the field.

As demonstrated in Fig.~\ref{fig_simualtions}a, the latent SDE model can better explain the data variance when using fewer latent dimensions compared to a latent ODE model. Increasing the latent dimensions beyond the true state dimensions does not significantly improve the model fit unlike the ODE-based model which stagnates at a lower absolute value and higher latent dimension.
Additionally, as shown in Fig.~\ref{fig_simualtions}b, the latent SDE model is more robust to process noise compared to latent ODE and LFADS models. These results suggest that latent SDEs can be more reliable when modeling real-world data when the true state dimension is unknown and process noise can emerge from unobserved interactions. In this setting, a probabilistic initial state with deterministic dynamics, as in the case of latent ODEs and LFADS, may not be sufficient. Figures~\ref{fig_simualtions}c and~\ref{fig_simualtions}d show the model predictions for the spike counts and the underlying firing rates, respectively. Figure~\ref{fig_simualtions}e demonstrates the phase portrait of the inferred dynamical system compared to the ground truth simulated dynamical system.

\begin{figure*}[!t]
    \centering
    \includegraphics[width=1.0\textwidth]{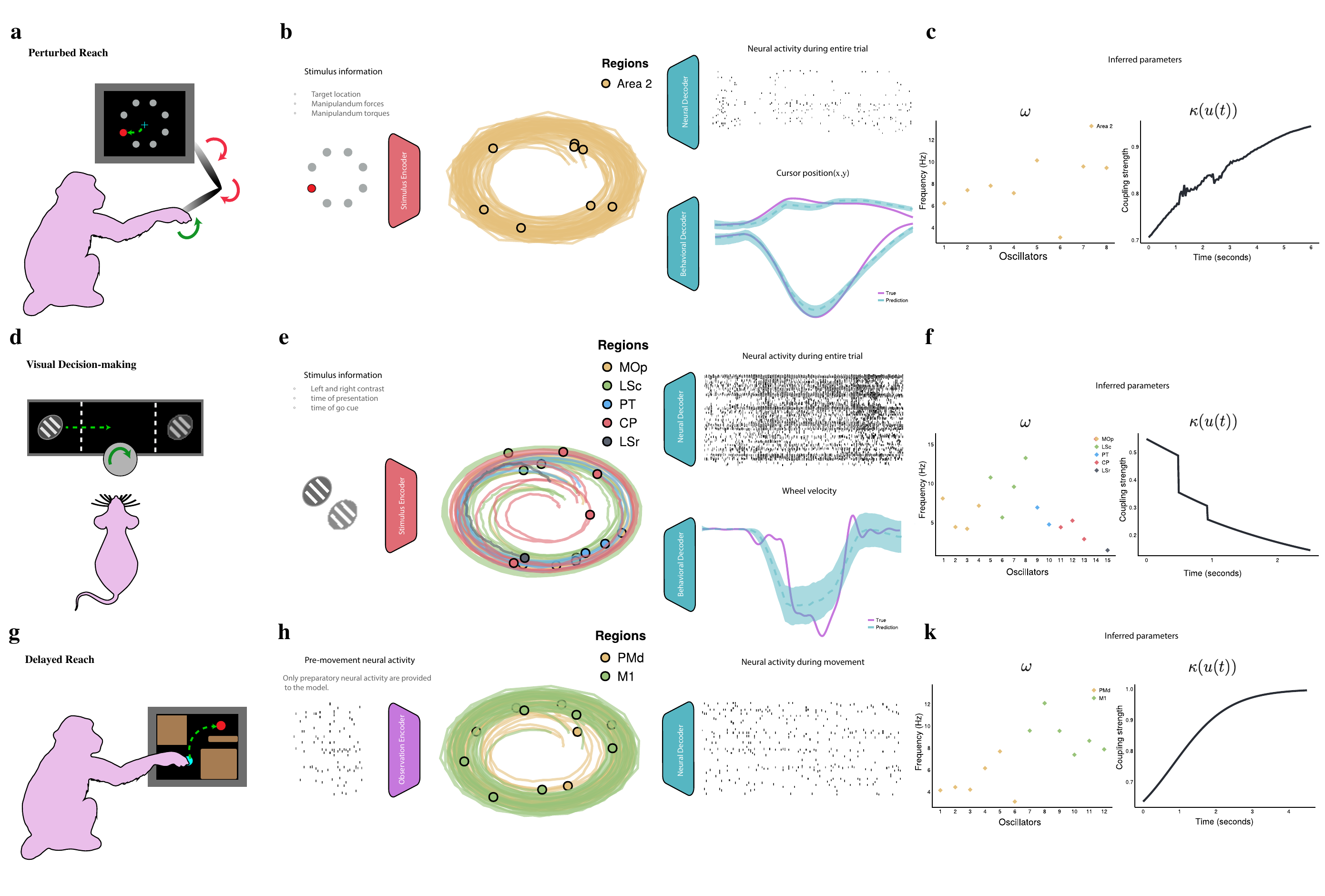}
    \caption{\textbf{Generative modeling of neural and behavioral data via latent coupled oscillators across three different datasets.} (\textbf{a-c}) \textbf{Perturbed reach task:} The generative model takes as input the target location and manipulandum forces/torques and is tasked with predicting neural responses of 64 Neurons in Area 2 and the behavioral response of the monkey as measured via the cursor position. The inferred frequency of the coupled oscillators as well as the coupling strength (for a sample trial) for one training run is show in \textbf{c}. (\textbf{d-f}) \textbf{Visual decision-making task:} Input consists of contrast levels as well as the timing of presentation and the go cue. The model is trained to predict both neural activity across multiple brain regions and wheel velocity. (\textbf{g,h,k}) \textbf{Delayed reach task:} Unlike the other tasks, this model receives only preparatory neural activity before movement onset and is trained to predict subsequent neural activity in the dorsal pre-motor (PMd) and primary motor cortex (M1) during movement execution. Abbreviations: 
    MOp: primary motor area, LSc: lateral sensory cortex, PT: posterior thalamus, CP: caudoputamen, LSr: Lateral sensory  rostral area, PMd: dorsal pre-motor cortex, M1: primary motor cortex. Monkey and mouse illustrations adapted from \url{scidraw.io} (CC-BY 4.0).}
    \label{fig_empirical}
\end{figure*}

\subsection*{Validation on empirical data}

\begin{table*}[!t]
\caption{Performance comparison of latent sequential models across different tasks and metrics. Prediction of neural responses is evaluated via computing the bits per spike (bps), and behavioral responses evaluated via $R^2$. The number of parameters of the generative dynamics for each model is shown in average across experiments. Encoders and decoder architectures are shared across all models. Arrows (↑/↓) indicate whether higher or lower values are better. Best performances are highlighted in \textbf{bold}. The reported results are the mean and std. of 5-fold cross validation.}
\vspace{10pt}
\centering
\begin{small}
\resizebox{\textwidth}{!}{
\begin{tabular}{lrlllll}
\hline
& & \multicolumn{5}{c}{\textbf{Datasets}} \\
\cmidrule{3-7}
\textbf{Latent Dynamics} & \textbf{\# Params} ↓ & \multicolumn{2}{c}{\textbf{Perturbed Reach}} & \multicolumn{2}{c}{\textbf{Visual Decision}} & \textbf{Delayed Reach} \\
& & Neural & Behavior & Neural & Behavior & Neural \\
& & (bps) ↑ & ($R^2$) ↑ & (bps) ↑ & ($R^2$) ↑ & (bps) ↑ \\
\hline
RNN & 8320 & 0.175 {\scriptsize ± 0.008} & 0.84 {\scriptsize ± 0.02} & 0.14 {\scriptsize ± 0.015} & 0.59 {\scriptsize ± 0.03} & 0.291 {\scriptsize ± 0.012} \\
GRU & 24960 & 0.187 {\scriptsize ± 0.004} & 0.81 {\scriptsize ± 0.015} & 0.16 {\scriptsize ± 0.008} & 0.63 {\scriptsize ± 0.02} & 0.305 {\scriptsize ± 0.006} \\
LSTM & 33280 & 0.191 {\scriptsize ± 0.003} & 0.85 {\scriptsize ± 0.01} & 0.16 {\scriptsize ± 0.006} & 0.66 {\scriptsize ± 0.015} & 0.309 {\scriptsize ± 0.004} \\
Neural ODE & 4880 & 0.221 {\scriptsize ± 0.005} & \textbf{0.86 {\scriptsize ± 0.012}} & 0.15 {\scriptsize ± 0.01} & 0.59 {\scriptsize ± 0.025} & 0.316 {\scriptsize ± 0.008} \\
CO-ODE & 465 & 0.221 {\scriptsize ± 0.006} & 0.85 {\scriptsize ± 0.015} & 0.21 {\scriptsize ± 0.009} & 0.61 {\scriptsize ± 0.02} & 0.309 {\scriptsize ± 0.007} \\
GP-SLDS & NA & 0.201 {\scriptsize ± 0.004} & NA & 0.08 {\scriptsize ± 0.01} & NA & \textbf{0.322 {\scriptsize ± 0.004}} \\
\hline
Linear SDE & 304 & 0.147 {\scriptsize ± 0.01} & 0.81 {\scriptsize ± 0.02} & 0.18 {\scriptsize ± 0.012} & 0.63 {\scriptsize ± 0.02} & 0.207 {\scriptsize ± 0.01} \\
Neural SDE & 4944 & \textbf{0.243 {\scriptsize ± 0.007}} & \textbf{0.86 {\scriptsize ± 0.015}} & 0.18 {\scriptsize ± 0.012} & \textbf{0.70 {\scriptsize ± 0.02}} & 0.317 {\scriptsize ± 0.009} \\
CO-SDE & 526 & 0.238 {\scriptsize ± 0.008} & \textbf{0.86 {\scriptsize ± 0.018}} & \textbf{0.22 {\scriptsize ± 0.011}} & 0.67 {\scriptsize ± 0.025} &  0.314 {\scriptsize ± 0.01}\\
\hline
\end{tabular}
}
\end{small}
\label{table1}
\end{table*}

\subsubsection*{Datasets \& baselines}
We evaluated our framework across three neuroscience datasets spanning different brain regions and behavioral tasks.

\paragraph{Perturbed reach task} In this experiment, a monkey is presented with a target location and tasked with moving the cursor from the center to the target location by controlling a manipulandum~\cite{chowdhury2020area} (Fig.~\ref{fig_empirical}a). On one portion of the trials, the monkey’s arm was perturbed before the reach by applying resisting forces on the manipulandum. Neural activity was recorded from Brodmann’s area 2 of the somatosensory cortex. Here, we trained a model to predict stimulus-evoked neural responses in area 2 as well as the hand position of the monkey. The input to the model are the target location, as well as the perturbing force and torques on the manipulandum. At test-time, we provide the first 50 ms of observations before stimulus presentation to the observation encoder to identify the initial condition.

\paragraph{Visual decision making}  In this task, a mouse is presented with images of different contrasts and the mouse is tasked with turning a wheel to bring the image with higher contrast to the center of the screen when a go cue is randomly presented~\cite{steinmetz2019distributed} (Fig.~\ref{fig_empirical}d). Neural activity is recorded in multiple regions across the brain using Neuropixels probes. Here, we trained the model to predict stimulus-invoked neural responses across several brain regions simultaneously as well as predict the wheel velocity. The input to the model are the contrast levels, the timing of their presentation, and the timing of the go cue. At test-time, we provide the first 50 ms of observations before stimulus presentation to the observation encoder to identify the initial conditions.

\paragraph{Delayed reach task} The task is a delayed center-out reach task with barriers, resulting in a variety of straight and curved trajectories~\cite{mante2013context} (Fig.~\ref{fig_empirical}g). Neural activity was recorded from the dorsal pre-motor (PmD) and primary motor (M1) cortices, and cursor, monkey gaze position, and monkey hand position and velocity are also provided. Here we train a model to predict neural responses during movement in both areas given the preparatory activity before the go cue. No stimulus information is provided to the model during training or prediction. At test-time, we provide the first 100 ms of neural observations before the go cue to the observation encoder to identify the initial conditions.

We compare different latent variable models across the three datasets. To enable a fair comparison, we fix the encoder and decoder architectures across all models while varying the latent sequential model. That is, we compare the performance of RNN, GRU, LSTM, Neural ODE, and a coupled oscillator ODE (CO-ODE) models with that of our proposed CO-SDE and a Neural SDE model. We also include a linear SDE model which employs linear drift and diffusion functions. All the baseline models had a probabilistic initial state and were trained via variational inference. Note that this formulation enables generalizing the baselines to several previous models proposed in literature~\cite{pandarinath2018inferring, hurwitz2021targeted, kim2021inferring, versteeg2023expressive}. We also compare against a recently proposed Gaussian process switching linear dynamical system (GP-SLDS)~\cite{hu2024modeling} (See Supplementary Information for details on the baselines).

\subsubsection*{Empirical results}
Table~\ref{table1} presents 5-fold cross-validation results across the three datasets, evaluating both neural activity prediction (in bits per spike) and behavioral performance (via the coefficient of determination, $R^2$) (see Supplementary Information for further details on the evaluation metrics).
First, we observe that our stochastic models (Neural SDE and CO-SDE) on average consistently outperform their deterministic counterparts. This suggests that incorporating stochasticity into the latent dynamics is crucial for capturing the variability inherent in neural and behavioral data and that a stochastic initial state might not be sufficient to capture uncertainty in the dynamics.

Second, we find that non-linear latent dynamics are crucial for accurate joint prediction of neural and behavioral responses. This is suggested by the performance of the linear SDE baseline which performs competitively to non-linear baselines only on predicting behavioral responses on the perturbed reach task. However, its performance is significantly lower for predicting neural responses across all three tasks. This suggests that while linear dynamics may suffice for capturing certain behavioral regularities, nonlinear models are necessary for accurate neural population modeling at the level of single-neuron. 

Third, we observe that model complexity, as measured by parameter count, does not correlate with performance. The coupled oscillator models achieve competitive or superior performance while using only 465-526 parameters, compared to 33,280 parameters for the LSTM model. Note that while this number does not account for the number of parameters required for modeling the approximate posterior SDE (3528), this remains a significant improvement in parameter efficiency for the generative model. This suggests that incorporating appropriate inductive biases through the coupled oscillator architecture enables more efficient modeling of the underlying dynamics. This also suggests that  standard deep learning models may be too over-parameterized to learn the underlying dynamical system during simple behavioral tasks.

Beyond quantitative performance, oscillator-based models offer interpretable parameterizations that connect to established neuroscience concepts. We visualize the inferred natural frequencies and coupling strengths learned by the model in Fig.~\ref{fig_empirical}. In both motor tasks -- the perturbed reach task with recordings from somatosensory cortex and the delayed reach task with recordings from two motor cortical areas -- we observe that coupling strength increases over time during movement preparation and execution (Fig.\ref{fig_empirical}c,k). This pattern is consistent with theories of neural communication through enhanced synchronization~\cite{fries2015rhythms}, where increasing oscillatory coupling may facilitate information integration across neural populations during coordinated motor output.
In contrast, the visual decision-making task shows a markedly different pattern (Fig.\ref{fig_empirical}f). Here, coupling strength decreases over time, with two distinct drops coinciding with the stimulus presentation and go cue. This temporal modulation may reflect dynamic routing of information\cite{fries2015rhythms}, where coupling is selectively reduced to gate irrelevant signals or reallocate processing resources as task demands shift from sensory encoding to decision formation and action selection.
We also observe systematic differences in natural frequencies across brain regions. Notably, oscillators corresponding to M1 exhibit higher average frequencies than those in PMd during the delayed reach task (Fig.\ref{fig_empirical}h). This frequency gradient is consistent with the hierarchical organization of motor cortex, where M1 operates at faster timescales than PMd, which is more involved in movement planning~\cite{Thura2022, churchland2012neural}. Such frequency differences may reflect distinct computational roles: higher frequencies in M1 may support rapid sensorimotor transformations required for real-time motor control, while lower frequencies in PMd may be suited for integrating information over longer timescales during motor preparation.
Important note on interpretation: While these learned parameters show relative consistency across model initializations and align with neuroscience literature, we emphasize that our method does not provide theoretical guarantees for parameter identifiability—a well-known challenge in nonlinear dynamical systems~\cite{miao2011identifiability}. Therefore, we present these findings as hypothesis-generating observations that suggest our model captures functionally meaningful structure, rather than as definitive claims about the underlying neural mechanisms. The value of these parameterizations lies in providing neuroscience-motivated, interpretable descriptions of population dynamics that can guide future experimental investigations.

\section*{Discussion}
This work introduces a probabilistic framework for modeling neural dynamics that combines the expressiveness of neural networks with the interpretability of mechanistic or phenomenological models. Our empirical results across simulated and real datasets demonstrate several key advantages of this approach.

First, explicitly modeling probabilistic dynamics through SDEs consistently improves model performance compared to deterministic alternatives, even when the latter incorporate uncertainty through probabilistic initial conditions. This suggests that neural variability stems not just from uncertain initial states but from inherently stochastic dynamics~\cite{laing2009stochastic}. The superior performance of SDE-based models in high-noise regimes as shown in the simulations results further supports this conclusion. Additional simulation results also show that using stochastic dynamics can improve model fit while relying on lower-dimensional latent states; an observation that is also found in stochastic low-rank RNNs~\cite{pals2024inferring}.

Second, our hybrid coupled oscillator model achieves competitive or superior performance while requiring an order of magnitude fewer parameters than standard deep learning approaches. This parameter efficiency suggests that incorporating appropriate inductive biases through mechanistic components can effectively constrain the model space~\cite{rackauckas2020universal, elgazzar2024universal}. The success of coupled oscillators in particular hints at their potential as a natural description language for neural population dynamics, consistent with existing theoretical work on neural oscillations~\cite{buzsaki2004neuronal, breakspear2010generative, churchland2012neural,bick2020understanding}. This also supports the potential of coupled oscillators as an expressive and efficient tool to model arbitrary time-series data, as seen in emerging machine learning research~\cite{rusch2020coupled, effenberger2022biology,rusch2024oscillatory}. Future work could explore in depth the connection between features of inferred coupled oscillators (e.g. synchronization, phase-amplitude coupling, etc.) and observed neural and behavioral responses. 

While we focus on coupled oscillator models motivated by motor cortex physiology, our framework readily accommodates diverse dynamical systems including Van der Pol oscillators~\cite{van1926lxxxviii} for modeling limit cycles with different relaxation properties, FitzHugh-Nagumo~\cite{fitzhugh1961impulses} models for excitable systems, attractor-based models for decision-making circuits~\cite{wang2002probabilistic}, or predator-prey dynamics for competitive interactions~\cite{galadi2021capturing}. The choice of dynamical model should be guided by domain knowledge and scientific hypotheses about the neural system. This flexibility enables researchers to test specific mechanistic hypotheses while maintaining the benefits of probabilistic inference and uncertainty quantification. Future work could systematically compare these alternative models on the same datasets to identify which dynamical structures best capture different neural computations. Alternatively, the  symbolic expressions that define the latent dynamics could be inferred using black-box optimization techniques such as differential genetic programming~\cite{VanGerven2025Neuromorphic}.

With that said, there are a number of limitations to the proposed framework which merits consideration. First, similar to most non-linear latent variable models in the literature, our approach does not guarantee parameter identifiability. Rather, our approach enables interpreting the system dynamics as a function of the behavior of the proposed system dynamics. Further work is required to ensure identifiability of the parameters of the latent dynamical system to enable robust comparison of different models~\cite{dunker2014parameter, hasan2021identifying}. Here, advances in epistemic AI that focus on quantifying parameter uncertainty are of particular relevance~\cite{Manchingal2025Epistemic}.
Another limitation is that our framework relies on variational inference for training the model parameters. This could be suboptimal as variational inference is known to provide overconfident uncertainty estimates and is prone to several optimization challenges such as posterior collapse, especially when coupled with powerful decoders~\cite{blei2017variational}.  That is, the model can rely on expressive decoders to model the data, rendering latent variables redundant. A potential solution is to utilize invertible decoders~\cite{zhou2020learning, versteeg2023expressive} to guarantee that the model relies on latent dynamics to model neural and behavioral observations.  Furthermore, training the model parameters currently requires back-propagation through an SDE solver, which can be computationally expensive in terms of both time and memory requirements. Potential solutions could lie in emerging simulation-free approaches for training (neural) stochastic differential equations~\cite{course2024amortized, zhang2024trajectory, bartosh2025sde}.

In this work, we provided a first demonstration of the usefulness of developing biophysics-inspired latent SDE models. 
Future work could explore comparing different dynamical systems beyond coupled oscillators and focus on extending the current approach to model neural data from several subjects as well as other recording modalities. The use of these models for neural system identification may provide new insights into neural information processing. Furthermore, their interpretability and expressiveness may pave the way for new applications in clinical neuroscience.

\section*{Acknowledgements}

This publication is part of the project Dutch Brain Interface Initiative (DBI$^2$) with project number 024.005.022 of the research programme Gravitation which is (partly) financed by the Dutch Research Council (NWO).

\bibstyle{plos2015}
\bibliography{main}

\clearpage
\section*{Supplementary Information}

\section{Learning the parameters of the generative model and inferring latent paths}\label{apdx_vi}
Recall that our goal is to learn a generative model $p_\theta(y,b \mid u)$ via the following continuous-time state-space model:
\begin{align}
&\dd x(t) = \mu_{\theta}(x(t), u(t))\dd t + \sigma_{\theta}(x(t), u(t)) \dd W(t), \quad t \in [0, \tau], \quad x_0 \sim \mathcal{N}(0, I)\\
&y(t) \sim p(y(t) \mid \lambda_\theta(x(t))), \quad b(t) \sim p(b(t) \mid \rho_\theta(x(t))) \,.
\end{align}
We leverage variational inference to train the model parameters $\theta$ in a tractable and scalable manner. The SDE induces a path measure $\mathcal{P}\tau$, which together with the initial distribution implicitly define our prior distribution $p_\theta(x \mid u)$ over continuous paths.
To define the posterior distribution, we consider another continuous-time stochastic process defined via the following SDE:
\begin{equation}
\dd x(t) = \nu_{\phi}(x(t), y(t), b(t), u(t))\dd t + \sigma_{\theta}(x(t), u(t)) \dd W(t), \quad t \in [0, \tau], \quad x_0 \sim \mathcal{N}(\alpha_{\phi}(y,b), \beta_{\phi}(y,b))
\end{equation}
This SDE induces a measure on the path space $\mathcal{Q}\tau$ which together with the initial distribution implicitly define our posterior distribution $q_\phi(x \mid u, y, b)$ over continuous paths.
The choice of using a similar diffusion function allows us to derive a tractable relationship between the two path measures induced via the two SDEs using Girsanov's theorem. According to Girsanov's theorem, the Radon-Nikodym derivative between these measures is given by:
\begin{equation}
\frac{d\mathcal{Q}\tau}{d\mathcal{P}\tau} = \exp\left(-\frac{1}{2}\int_0^\tau |\Delta(t)|^2 \dd t + \int_0^\tau \Delta(t)^T \dd W(t)\right)
\end{equation}
where $\Delta(t) = \sigma_{\theta}(x(t), u(t))^{-1}(\nu_{\phi}(x(t), y(t), b(t), u(t)) - \mu_{\theta}(x(t), u(t)))$.
We can thus derive a finite Kullback–Leibler divergence between the two path measures as:
\begin{align}
D_\text{KL}(\mathcal{Q}\tau || \mathcal{P}\tau) = \frac{1}{2}\mathbb{E}_{\mathcal{Q}\tau}\left[\int_0^\tau |\Delta(t)|^2 \dd t\right]
\end{align}
The evidence lower bound (ELBO) can then be written as:
\begin{align}
\text{ELBO}(\theta, \phi) 
&= \mathbb{E}_{q_\phi(x \mid u,y,b)}\left[\log p_\theta(y,b \mid x,u)\right] - D_\text{KL}(\mathcal{Q}\tau || \mathcal{P}\tau) - D_\text{KL}(\mathcal{N}(\alpha_\phi(y,b), \beta_\phi(y,b)) || \mathcal{N}(0, I))\\
&= \mathbb{E}_{q_\phi(x \mid u,y,b)}\left[\int_0^\tau \log p_\theta(y(t),b(t) \mid x(t),u(t))\dd t\right] 
- \frac{1}{2}\mathbb{E}_{\mathcal{Q}\tau}\left[\int_0^\tau |\Delta(t)|^2 \dd t\right] \\
&\quad - \frac{1}{2}\left(\text{tr}(\beta_\phi(y,b)) + |\alpha_\phi(y,b)|^2 - \log\det(\beta_\phi(y,b)) - d_x\right)
\end{align}
This optimization can be performed using stochastic gradient descent, where the path integrals are approximated using numerical integration schemes such as the Euler-Maruyama method, and the expectations are estimated using Monte Carlo sampling.

\section{Implementation details}\label{apdx_impl}
In the following, we summarize the general implementation details of our latent SDE.
The models were trained to minimize ELBO on the training set by standard stochastic gradient backprogation through the computational graph. This includes the numerical solver. That is, the intermediate steps are saved in memory during the forward pass and used to update the parameters during the backward pass. While this scales to $\mathcal{O}(n)$ in terms of memory requirement where $n$ is the number of steps in the numerical solver compared to $\mathcal{O}(1)$ of the adjoint method, we found that this leads to more stable training in our experiments. We used an Euler-Maruyama solver for all our experiments. We found that setting a step size $\dd t$ at 2 times the bin width of the spikes in our empirical experiments did not significantly influence the performance and helped improved the training speed and memory requirements. We ran a hyperparameter search for the latent states, which in our oscillatory models (CO-ODE, CO-SDE) represents the number of oscillators $\times 2 $. We found that the optimal number of oscillators varies across tasks (typically between 8 and 16) for the evalutaed tasks.
We also found that cyclic annealing the KL-divergence term in our ELBO significantly improved training stability. We use a linear with 4 cycles with a linearly increasing value from 0 to 10 for half the cycle and then stays constant for the second half, until the cycle resets. In experiments which include reconstructing both neural and behavioral data, we use a constant weighting for each during training depending on the dataset. Our experiments were carried out on a  NVIDIA A100-PCIE-40GB. Table~\ref{table2} lists the main components of our latent SDE and their hyperparameters used across all empirical experiments. 

\begin{table}[htbp]
\caption{General hyperparameters and configuration settings used during training.}
\vspace{10pt}
\centering
\begin{small}
\begin{tabular}{llr}
\toprule
\textbf{Component} & \textbf{Hyperparameter} & \textbf{Value} \\ 
\midrule
\multirow{8}{*}{\textbf{General}} 
 & Batch size & 128 \\
 & Optimizer & Adam \\
 & Learning rate & $10^{-3}$ \\
 & Max epochs & 2000 \\
 & Patience & 20 \\
 & MC samples (training) & 1 \\
 & MC samples (prediction) & 30 \\
 & KL annealing scheduler & Cyclic \\ 
\midrule
\multirow{4}{*}{\textbf{\shortstack[l]{Differential equation\\solver}}} 
 & Solver & Euler-Maruyama \\
 & Step size & $2(t_{i+1}-t_i)$ \\
 & \multirow{2}{*}{Backpropagation} & Discrete-adjoint \\
 & & (discretize-then-optimize) \\
\bottomrule
\end{tabular}
\label{table2}
\end{small}
\end{table}

\section{Architecture}\label{apdx_arch}
We summarize below the objective and parameters of each module in our proposed framework, detailing the configurations used in our empirical experiments.
\paragraph{Observation Encoder}
The observation encoder serves two purposes: (1) determining the parameters of the posterior distributions for the initial state, and (2) generating a lower-dimensional time-series representation (context) to condition the augmented SDE on the data. This is accomplished through two parallel components: the initialization network and the context network.
The initialization network processes the first 30\% of the observations using a 2-layer LSTM with hidden size 64, applying it in reverse order. The network then transforms the output through two parallel affine transformations into the latent dimension size, with a soft-plus activation applied to the variance component. This generates the mean and squared variance of the posterior initial condition.
The context network processes the complete time-series of observations through a 2-layer LSTM, producing a time-series of length 64. This context vector undergoes linear interpolation to create a continuous-time representation for the augmented SDE.
\paragraph{Stimulus Encoder}
This module creates a lower-dimensional continuous representation of the stimulus. For low-dimensional stimuli (where $d_v \leq d_x$), we employ an identity transformation. We include time as an additional input channel. In scenarios without stimulus information (such as the maze reach experiment), we set $u(t) = t$. After encoding, we use linear interpolation to obtain a continuous-time representation of the stimulus.
\paragraph{Generative SDE}
The generative SDE defines our dynamics by mapping the current state and encoded stimulus to the change in the next state. The drift and diffusion structure is flexible and should incorporate prior knowledge when available. In this paper, we present a model combining coupled oscillators with neural networks to represent the SDE's drift vector field. We also evaluate a neural SDE variant with a 2-layer MLP with hidden layer size 64 and tanh activation functions. The diffusion uses a sparse connection layer with non-zero diagonal parameters only, which is shared with the augmented SDE.
\paragraph{Augmented SDE}
The augmented SDE induces a path measure on continuous latent paths, implicitly defining our approximate posterior distribution. This neural SDE processes a concatenated vector of the state, encoded stimulus, and context at time $t$ to determine the change in the next state. Our implementation uses a 2-layer MLP with hidden size 64 and tanh activation functions for the drift vector field. The diffusion function is identical to that of the generative SDE.
\paragraph{Decoder}
The decoder transforms latent states into observations. For systems with both neural and behavioral observations, we implement parallel neural and behavioral decoders. The neural decoder consists of an MLP with one hidden layer of size 128 and soft-plus activations. For $K$ neural populations, we employ $K$ similar MLPs, each with output size matching the neuron count in its corresponding population. These outputs' exponents define the rate parameter of a time-inhomogeneous Poisson distribution.
The behavioral decoder uses two parallel linear transformations on the latent states to generate the mean and squared variance of a Gaussian distribution at each time step, defining the predicted behavior distribution.

\section{Baseline Model Specifications}\label{apdx_baselines}

To ensure a fair comparison across all models, we fix the encoder and decoder architectures while varying only the latent sequential dynamics model. This section provides comprehensive architectural details and parameter counts for each model variant.

\subsection*{Training Configuration}

All models (except gpSLDS) were trained using the hyperparameters specified in Table~\ref{table2}. Key training details include:
\begin{itemize}
    \item All models trained via variational inference with ELBO objective
    \item Probabilistic initial state: $x_0 \sim \mathcal{N}(\mu_{\phi}(y, b), \Sigma_{\phi}(y, b))$ where $\mu_{\phi}$ and $\Sigma_{\phi}$ are produced by the shared observation encoder
    \item For SDE models: Euler-Maruyama solver with step size $\dd t = 2 \times \text{bin width}$
    \item For ODE models: Euler solver with same step size
    \item For RNN/GRU/LSTM: Standard discrete-time updates at observation time points
    \item Cyclic KL annealing applied to all variational models
\end{itemize}

\subsection*{Implementation Notes}

\paragraph{Neural ODE \& Neural SDE}
These models use flexible neural network parameterizations for the drift (and diffusion) functions. The MLP architecture allows the model to learn arbitrary smooth dynamics while maintaining computational tractability through modern ODE/SDE solvers.

\paragraph{Coupled Oscillator Models (CO-ODE \& CO-SDE)}
The coupled oscillator models incorporate mechanistic priors about neural oscillations. Each oscillator evolves according to:
\begin{equation}
\frac{\dd^2 x_i}{\dd t^2} + \gamma_i \frac{\dd x_i}{\dd t} + \omega_i^2 x_i = \sum_{j \neq i} c_{ij}(u(t)) x_j
\end{equation}
where the coupling strengths $c_{ij}(u(t))$ are modulated by the stimulus through a small neural network. This inductive bias significantly reduces the parameter count while maintaining expressiveness.

\paragraph{Recurrent Models (RNN, GRU, LSTM)}
These discrete-time models are evaluated at observation time points. To ensure fair comparison with continuous-time models, we:
\begin{itemize}
    \item Sample hidden states at the same time points as ODE/SDE models
    \item Use the same batch size and training duration
    \item Apply the same variational inference framework with probabilistic initial states
\end{itemize}

\paragraph{Fair Comparison Justification}
By fixing encoder and decoder architectures, we isolate the impact of different latent dynamics models on performance. Any differences in results can be attributed solely to how well each model captures the temporal dynamics of neural and behavioral data. The parameter counts in Table 1 of the main text reflect only the latent dynamics parameters, as the encoder/decoder parameters (shared across all models) do not contribute to performance differences.

\section*{Evaluation Metrics}

\subsection*{Bits per Spike}

We assess the quality of neural response predictions using the \textit{bits per spike} (bps) metric, an information-theoretic quantity that measures the gain in predictive log-likelihood of a model over a time-homogeneous Poisson baseline, normalized by the number of spikes. The metric is grounded in the Poisson log-likelihood and has been widely adopted in recent neural latent variable modeling literature.

Let $y_{i,t} \in \mathbb{N}$ denote the observed spike count for neuron $i$ at time bin $t$, and let $\hat{\lambda}_{i,t}$ denote the firing rate predicted by the model. Let $\bar{\lambda}_i$ denote the empirical mean firing rate of neuron $i$ over the evaluation data. The bits per spike is computed as:
\begin{equation}
\text{bps} = \frac{1}{\sum_{i,t} y_{i,t}} \sum_{i,t} \left[ y_{i,t} \log_2 \left( \frac{\hat{\lambda}_{i,t}}{\bar{\lambda}_i} \right) - (\hat{\lambda}_{i,t} - \bar{\lambda}_i) \log_2 e \right] \,.
\end{equation}

This expression reflects the difference in log-likelihood (in bits) between the model and a baseline that predicts a constant firing rate for each neuron. Positive values indicate better predictive performance than the baseline; zero corresponds to parity; negative values indicate worse-than-baseline predictions.

Our use of this metric is inspired by~\cite{pei2021}, who introduced bits per spike as a normalized log-likelihood score in the context of \textit{co-smoothing}, wherein the goal is to predict the activity of held-out neurons conditioned on held-in neurons on the same trial. In contrast, we apply this metric to \textit{all recorded neurons} on entirely \textit{leave-out trials} that are not seen during training. This choice evaluates the model’s ability to generalize temporally, rather than across neurons, and aligns with our objective of modeling neural dynamics over continuous time.

\subsection*{Coefficient of Determination ($R^2$)}

To evaluate the model’s ability to predict behavioral measurements (e.g., kinematic variables or choice-related signals), we use the coefficient of determination, $R^2$, under a Gaussian readout from the inferred latent state. This metric measures the proportion of variance in the observed behavior explained by the model's predictions.

Let $y_t \in \mathbb{R}^d$ be the observed behavioral measurement at time $t$, and $\hat{y}_t$ the corresponding model prediction. The $R^2$ score is defined as:
\begin{equation}
R^2 = 1 - \frac{\sum_t \| y_t - \hat{y}_t \|^2}{\sum_t \| y_t - \bar{y} \|^2}
\end{equation}
where $\bar{y}$ is the empirical mean of $y_t$ over the evaluation set. An $R^2$ score of 1 indicates perfect prediction, 0 corresponds to performance equal to predicting the mean, and negative values reflect worse-than-mean predictions. This metric provides an interpretable measure of behavioral predictability from latent neural dynamics.

\section*{Supplementary References}

\end{document}